\documentclass[11pt,preprint]{aastex}

\begin{document}

%

\title{Light Propagation in Inhomogeneous Universes.\\
IV. Strong Lensing and Environmental Effects.}

\author{Premana Premadi\altaffilmark{1} and Hugo Martel\altaffilmark{2,3}} 

\altaffiltext{1}{Department of Astronomy and Bosscha Observatory,
Bandung Institute of Technology, Bandung, Indonesia}

\altaffiltext{2}{Department of Astronomy, University of Texas, Austin, 
                 TX 78712}
\altaffiltext{3}{D\'epartement de physique, de g\'enie physique et d'optique,
Universit\'e Laval, Qu\'ebec, Canada, G1K 7P4}

\begin{abstract}
We study the gravitational lensing of high-redshift sources in a $\Lambda$CDM
universe. We have performed a series of ray-tracing experiments, and
selected a subsample of cases of strong lensing (multiple images, arcs, and
Einstein rings). For each case, we identified a massive
galaxy that is primarily responsible for
lensing, and studied how the various density inhomogeneities along the line
of sight (other galaxies, background matter) affect the properties of the
image. The matter located near the lensing galaxy, and physically associated
with it, has a small effect. The background matter increases the magnification
by a few percents at most, while nearby galaxies can increase it by up to about
10 percent. The effect on the image separation is even smaller.
The only significant effect results from the random alignment of
physically unassociated galaxies, which can increase the magnification
by factors of several, create additional images, and turn arcs into rings.
We conclude that the effect of environment on strong lensing is
negligible in general, and might be important only in rare cases. 
We show that our
conclusion does not depend on the radial density profile
of the galaxies responsible for lensing.
\end{abstract}


\newpage
%

\section{INTRODUCTION}

The gravitational lensing of sources located at cosmological distances
is usually referred to as ``strong lensing'' or ``weak lensing,'' depending
on the magnitude of the effect.
``Strong lensing'' refers to cases
with multiple images, arcs, or rings, while ``weak lensing'' refers
to the magnification and shear of single images. We can divide the
observed cases of strong lensing in two groups. The first group contains 
the ``arc second'' cases: multiple-image systems with image separations
of order arc seconds, or rings with radii of that order
(see Kochanek et al. 1998).\footnote{See http://cfa-www.harvard.edu/castles.} 
In such cases, the lens is usually a
single, massive galaxy located along or near the line of
sight to the source.
A classic example is Q0957+561, the first gravitational lens
discovered (Walsh, Carswell, \& Weymann 1979). 
The second group contains the ``arc minute'' cases, in which the lens
is an entire cluster of galaxies. These lenses produce mostly giant arcs,
with radii in the range $15''-60''$ (see Table~1 of Williams, Navarro,
\& Bartelmann 1999). 
The most famous case is the cluster
CL~0024+1654, located at redshift $z=0.39$,
which produces three main arcs and one counter-arc, with a radius
of $35''$. Both strong and weak lensing have been used to constrain the
cosmological parameters, the spectrum of primordial density fluctuations,
and the nature and structure of the lenses (for recent reviews, see
Wambsganss 1998; Soucail 2001; Bartelmann \& Schneider 2001;
Claeskens \& Surdej 2002).

In this paper, the fourth of a series on light propagation
in inhomogeneous universes, we focus on the arc-second cases of
strong lensing that result when a massive galaxy is
located along the line of sight to a distant source. 
If the galaxy were the only density inhomogeneity along the
line of sight, a description of its lensing properties would be
quite trivial. There is a large body of literature describing
the lensing properties of individual galaxies (or more generally ``halos'')
with particular density profiles, including spherically symmetric
profiles (e.g. Hinshaw \& Krauss 1987; Schneider, Ehlers, \& Falco 
1992 [hereafter SEF], \S8.1.5; Wright \& Brainerd 2000;
Rusin \& Ma 2001; Martel \& Shapiro 2003) and flattened ones 
(e.g. Schramm 1990; Schneider \& Weiss 1991; Keeton \& Kochanek 1995).
However, treating lensing galaxies in isolation is only a
convenient approximation.
The presence of other density inhomogeneities along the same
line of sight can modify the properties of the image. If a lensing galaxies
is located in a dense cluster, the presence of nearby cluster members
can modify the magnification or the image
separation. Also, the whole cluster might be embedded inside a
common dark matter halo (hereafter referred to as
the {\it background component}), whose presence might also have
some effect of the lens properties. Strong lensing usually requires a massive
elliptical or S0 galaxy, and such galaxies tend to lie in dense
environments, as indicated by the morphology-density relation
(Dressler 1980). However, it has been suggested that
the luminosity distribution of galaxies depends on environment
(Bromley et al. 1998; Christlein 2000; Zabludoff \& Mulchaey 2000),
leading to a ratio of dwarf to giant galaxies that is higher in
denser environments. Keeton, Christlein, \& Zabludoff (2000) argue that
this effect nearly cancels the effect of the morphology-density
relation, and predict that only $\sim25\%$ of lensing galaxies are located
in dense environments. While environmental effects are irrelevant for the
remaining 75\% of lenses that do not reside in dense environments,
they can certainly be 
important for the 25\% of lenses located in such environments,
and might be necessary to explain the few cases with large angular
separations ($s>6''$) that have been observed (see, e.g., 
Inada et al. 2003).

Turner, Ostriker, \& Gott (1984, hereafter TOG) have considered
a simple model
in which lensing galaxies, modeled as
singular isothermal spheres,
are embedded into uniform sheets of background matter. They
showed that the effect of the background matter
can be important, but also indicated that this effect
can be weakened by the presence of low-density regions along
the same line of sight.
Treating the background matter as a uniform sheet allows an analytical 
treatment, leading to orders of magnitude estimates, but it is a rather
crude representation of the actual matter distribution in
a CDM universe in which large-scale structure forms by the growth of
Gaussian random fluctuations. In such universe, 
the rms density fluctuation $\sigma_{\rm rms}$ at a particular scale $\lambda$
is given by
\begin{equation}
\sigma_{\rm rms}^2(\lambda)
={1\over2\pi^2}\int_0^\infty P(k)W^2(k\lambda)k^2dk\,,
\end{equation}

\noindent where $P(k)$ is the power spectrum, and $W$ is the top-hat
filter function, given by
\begin{equation}
W(x)={3\over x^3}(\sin x-x\cos x)\,.
\end{equation}

\noindent
In a CDM universe, small scales collapse before large ones, hence there
is at any epoch a maximum scale $\lambda_{\max}$ corresponding
to the largest structures one expect to find at that epoch in a typical
region of the universe. To estimate $\lambda_{\max}$, we 
make the common assumption that the
largest structures found in a representative region of the universe
correspond to $3\sigma$ fluctuations. Then, according to the
Press-Schechter approximation, the radius of these structures is given
by solving
\begin{equation}
\sigma_{\rm rms}(\lambda_{\max})={\delta_{\rm crit}\over3}
\end{equation}

\noindent 
for $\lambda_{\max}$, where $\delta_{\rm crit}\approx1.69$. For
a $\Lambda$CDM universe, we get $\lambda_{\max}\approx23\,\rm Mpc$.
The typical diameter
of the largest objects (voids or clusters) we expect to find in
a representative region of the universe is therefore of
order $2\lambda_{\max}\approx46\,\rm Mpc$. Sources that are gravitationally
lensed are normally located at cosmological distances that greatly exceeds
that scale. Hence, there will be several overdensities and
underdensities along the line of sight to any source, and one could expect
a near-cancellation of their effects. TOG acknowledge that fact.

Two arguments can be made against this line of reasoning. First, while this
near-cancellation is expected for a typical line of sight, it might not
occur for an atypical one (this is the argument made by TOG). Second,
even though there might be several overdensities and underdensities
along the line of sight, their relative contributions to lensing
will differ.
It is well known that the matter that contributes most to lensing tends
to be located about half-way between the source and the observer.
To be more quantitative, consider a lens L with projected surface density
$\sigma$, located along the line of sight to a distant source S.
We define the convergence $\kappa$ of the lens as
\begin{equation}
\kappa={\sigma\over\sigma_{\rm crit}}\,,
\end{equation}

\noindent where $\sigma_{\rm crit}$ is the critical surface density,
defined by
\begin{equation}
\sigma_{\rm crit}={c^2D_S\over4\pi GD_LD_{LS}}\,,
\end{equation}

\noindent and $D_S$, $D_L$, and $D_{LS}$ are the angular diameter
distance between observer and source, observer and lens, and source
and lens, respectively (SEF, p.~158). 
The value of $\kappa$ provides an estimate of the
strength of gravitational lensing. We can define a {\it lensing weight}
$w(z)$ using
\begin{equation}
w(z)={1\over\sigma_{\rm crit}}={4\pi GD_LD_{LS}\over c^2D_S}\,.
\end{equation} 

\noindent The lensing weight gives a measure of the relative contributions to
lensing of matter located at various redshifts $z$, all
other things being equal. This function is zero at the location of the
observer and the source, and peaks at intermediate distances (see Fig.~2
below). The largest contribution to lensing will usually come from
matter located in the redshift interval where $w(z)$ is large.
The extent of this region is smaller than the whole distance between
source and observer. If it is only a few times $\lambda_{\max}$,
it will contain only a few
overdensities and underdensities, and the cancellation of their effect
will not be perfect in general.
Notice that this effect is partially compensated by the fact that at redshifts
$z>0$ where $w(z)$ peaks, 
the value of $\lambda_{\max}$ is smaller than at the present in a CDM
universe.

There is a clear need for a more realistic approach to this problem
than the uniform-sheet approximation of TOG.
A recent attempt to quantify the
importance of environmental effects on strong lensing by massive
galaxies was presented by Holder \& Schechter (2003). These authors
used simulated large-scale structure and galaxy distributions (as we do here),
identified the most massive galaxies as being the ones capable of producing
strong lensing, and calculated, at the location of these galaxies,
the gravitational shear caused by nearby large-scale structures and galaxies.
They concluded that the typical shear is of order $10-15\%$.

Our goal in this paper is similar. We want to determine the typical
magnitude of environmental effects, and demonstrate that
very strong effects are rare. However, we
consider an approach to this problem that drastically differs
from the one used in previous studies.
We use a multiple lens-plane algorithm to simulate the actual
gravitational lensing of a large number of distant sources.
We performed 100 simulations, each simulation producing the images of 841
sources, for a total of 84,100 images.
We extract from these simulations a subsample of images that
constitute examples of strong lensing, and we
reanalyze this subsample on a case-by-case basis.
While the original sample of
84,100 images is unbiased and statistically significant, the subsample of
16 cases is strongly biased, since it includes only cases of strong lensing.
Such biased sample is ideal for determining an upper limit to the effects
of environment on strong lensing.

This work differs from previous studies in our approach to
analyzing the effect of the density fluctuations along the line of sight.
In particular, instead of focusing on their cumulative effect,
we divide these fluctuations in different density components, and determine
their relative contributions independently.
The most prominent components are clusters of galaxies.
We divide these clusters into two components: the galaxies themselves
and the diffuse CDM halo in which the galaxies are embedded, and we study
the effects of these two components, both individually and in combination.
We also consider the effect resulting from the random alignment of
galaxies that are physically unassociated. 

The remainder of this paper is organized as follow: In \S2, we briefly
describe our numerical algorithm. In \S3, we describe our original
set of simulations, and the 16 cases of strong lensing that we have selected
for further analysis. In \S4, we identify, for each case, the particular
galaxy (referred to as {\it The Lens}) that is primarily responsible for
producing strong lensing. In \S5, we investigate the effects of the
various density components along the line of sight. In \S6, we experiment
with variations of the basic model, by considering different
density profiles for galaxies. 
Conclusions are presented in \S7.

\section{THE ALGORITHM}

We have developed a {\it multiple lens-plane algorithm} to study light 
propagation in inhomogeneous universes (Premadi, Martel, \& Matzner 1998,
hereafter Paper I;
Martel, Premadi, \& Matzner 2000; Premadi et al. 2001a, b). 
Here, we give a brief description of the algorithm. For more details,
we refer the reader to Paper I.

In this algorithm, the space between
the observer and the source is divided into a series of cubic boxes
of comoving size 128 Mpc, and the matter content of each box is projected
onto a plane normal to the line of sight. The trajectories of light
rays are then computed by adding successively the deflections caused by
each plane, using geometrical optics. This method is described in
SEF, chapter 9.

To use this algorithm, we need to provide a description of the
matter distribution along the line of sight. Matter is divided into two
components: background matter and galaxies. We use a $\rm P^3M$ algorithm
to simulate the distribution of background matter. The simulations used
$64^3$ equal-mass particles and a $128^3$ PM grid,
inside a comoving volume of size 128 Mpc. We then use a Monte Carlo
method for locating galaxies into the computational volume
according to the underlying distribution of background matter
(Martel, Premadi, \& Matzner 1998; Paper I). 
The distribution of galaxies satisfy several observational
constraints, including the two-point correlation function.
The luminosities and morphological types of galaxies are chosen
according to the Schechter luminosity function (Schechter 1976) and 
the morphology-density relation (Dressler 1980), respectively.
We treat galaxies as nonsingular isothermal spheres,
using the models of Jaroszy\'nski (1992). The projected surface density
profile of galaxies is given by
\begin{equation}
\sigma(r)=\cases{\displaystyle
{v^2\over4G(r^2+r_c^2)^{1/2}}\,,&$r\leq r_{\rm max}$;\cr
0\,,&$r>r_{\max}$;\cr}
\end{equation}

\noindent where $r$ is the projected distance from the
center, $v$ is the rotation velocity, $r_c$ is the core radius,
and $r_{\max}$ is a truncation radius introduced
to prevent the mass from diverging.
The parameters $v$, $r_c$, and $r_{\max}$ are
functions of the galaxy luminosity and morphological type, given by
\begin{eqnarray}
r_c&=&r_{c,0}(L/L_*)\,,\\
r_{\max}&=&r_{\max,0}(L/L_*)^{1/2}\,,\\
v&=&v_0(L/L_*)^\gamma\,,
\end{eqnarray}

\noindent where $L_*$ is the characteristic luminosity of the 
Schechter luminosity
function, and the parameters $r_{c,0}$, $r_{\max,0}$, $v_0$, and $\gamma$ are
given in Table~1. The values of $v_0$ and $\gamma$ are based on
the Faber-Jackson relation for early-type galaxies (Faber \& Jackson 1976;
de Vaucouleur \& Olson 1982) and the Tully-Fisher relation
for spiral galaxies (Tully \& Fisher 1988; Fukugita et al. 1991).
Equations~(8) and~(9) are based on the studies of Kormendy (1987)
and Holmberg (1973), respectively. We refer the reader to
Jaroszy\'nski (1992) for details.

\section{INITIAL SIMULATIONS}

We consider a flat, untilted,
COBE-normalized $\Lambda$CDM model with density parameter
$\Omega_0=0.3$, cosmological constant $\lambda_0=0.7$, Hubble constant
$H_0=65\,\rm km\,s^{-1}Mpc^{-1}$, 
baryon density parameter $\Omega_b=0.019h^{-2}$, and
CMB temperature $T_{\rm CMB}=2.73$. The CDM power spectrum of density 
fluctuations is described by the transfer function of Bardeen et al. (1986)
with the normalization of Bunn \& White (1997). These simulations are part
of the Texas P$^3$M database (Martel \& Matzner 2000; 
El-Ad et al. 2002), and are publicly available.
We performed a series of 100 ray-tracing experiments, each
experiment consisting of
propagating a square beam of angular size $21.9''\times21.9''$ containing
$341^2$ rays, from the observer to the source plane,
located at redshift $z_S=3$. These experiments altogether
produced images of 84,100 sources of angular diameter $1''$ located
at redshift $z_S=3$.
We selected for further analysis
a subsample of 16 images that represent cases of strong lensing.
These images are shown in Figure~1. Each case is labeled by a letter,
from A to P. Our sample includes double images (cases E, F, G, I, J, K),
triple images (cases A and C), an Einstein ring (case H), two-hole
rings (cases D and N), and more complex images resulting from
multiple deflections (Cases B, L, M, O, and P).

\section{IDENTIFYING THE LENS}

We assume that in each case selected, strong lensing results from
the presence of a massive galaxy along the line of sight. We shall refer to
this galaxy as {\it The Lens} (with capital initials). 
To identify it, we pick the ray closest to the
center of the source (the ``central ray''), 
and follow the trajectory of that ray from the source
to the observer, to find which galaxies are directly hit by this ray. 
If only one galaxy is hit, this galaxy is identified as The Lens.
If several galaxies are hit, we compute for each galaxy the convergence
$\kappa({\bf x})$ at the position $\bf x$ on the lens plane where
the ray hits the galaxy, using equation~(4).
The convergence provides a direct estimate of
the strength of gravitational lensing. The galaxy that produces
the largest value of the convergence
at the location of the central ray is identified as The Lens.

Figure 2 shows the masses and redshifts of
the galaxies being hit, for all 16 cases. For each case, the
red bar shows The Lens, while the blue bars show the other galaxies. 
For cases E and L, The Lens is the only galaxy being hit.
The solid curves show the lensing weight $w(z)\equiv1/\sigma_{\rm crit}$,
which measures
the relative contribution to lensing of matter located at various redshifts.
The lensing weight peaks in the region $z=[0.8,1.0]$, about half way between the
source and the observer. Not surprisingly, The Lens tends to be found in this
region, the only exception being for case M, where The Lens is much closer to
the source, at redshift $z_L=1.80$.  
The Lens is not always the most massive galaxy being hit. 
In particular, for cases
G, M, and N, a galaxy much more massive than The Lens was hit, but it was
located near the source and therefore had a small contribution. For case P,
two galaxies located at the same redshift are hit, and the least massive
is identified as The Lens, because that galaxy is hit near its center,
where the projected surface density $\sigma$ is the largest, while the other
galaxy is hit near its outskirt, where the projected surface density is small.

\section{EFFECT OF THE VARIOUS COMPONENTS}

\subsection{Images, Magnification, and Image Separation}

For each case, we identify four components that contribute or
might contribute to gravitational
lensing: (1) The Lens itself, (2) the galaxies located on the same lens 
plane, (3) the background matter located on the same
plane, and (4) the galaxies and background matter located on other planes.
Consecutive lens planes represent regions separated by a comoving distance
of 128~Mpc. This greatly exceeds the characteristic scale of both
the galaxy-galaxy and cluster-cluster 2-point correlation
functions, as well as the scale $\lambda_{\max}$ discussed in \S1. 
Hence, matter located on different lens planes is physically
unassociated, and this fact is built into the ray-tracing algorithm:
consecutive lens planes are generated by different N-body simulations,
and are given random shifts (Paper I). This is a standard technique to ensure
that the structures in neighboring planes are uncorrelated.

All these components were included in the original simulations described
in \S3.
We redid these simulations for our 16 selected cases, turning on only
certain components at a time, 
to estimate their effect. These simulations included (a) Lens only,
(b) Lens + galaxies on same plane, (c) Lens + background matter on same plane,
and (d) Lens + galaxies and background matter on same plane. The results are
shown in Figure 3, along with the results of our original ``all planes''
simulations. The striking result is that the effect of the background matter
and other galaxies on the same plane is very small. Except for cases B, K,
and P, it is hardly noticeable. Figure~4 shows the magnification $\mu$
for all cases. Differences between squares and triangles show the effect of the
background matter. Differences between open and filled symbols show the
effect of the galaxies. Differences between asterisks and filled squares
show the effect of the other lens planes.
Figure 5 shows the image separation $s$ for the 10 cases A, C--G, I--K, and N
that produce multiple images.
The separations are measured between the center of each image.
For cases with more than 2 images (such as case A), the separation is
measured between the two outermost images.
For cases D and N, no asterisks are plotted because these cases do not
produce multiple images when all planes are included.

From Figures~4 and~5, we can estimate the effect of the various
components. The effect of the background matter on the magnification is of
order 1\% at most, while the effect of the galaxies is of order 10\% at
most. The exceptions are cases L and M, for which the effect of galaxies
is less than 1\%, while the effect of the background matter is $\sim4\%$ for
case L and $\sim6\%$ for case M. The presence of galaxies near The Lens tends
to increase the magnification. However, for cases B, K, and N we find the 
opposite effect: the magnification is reduced.
The effect of these components on the image separation tends to be very small.
It exceeds 2\% for three cases only, D, F, and K. In all three cases,
it is the galaxies that are responsible for the effect. Notice that in cases D
and F, the effect is a reduction of the image separation.
The largest effect seen in Figures~4 and~5 result from the presence of
the other lens planes. In 6 cases out of 16 (A, C, D, M, N, and O), 
the magnification more than doubles when the other lenses planes are included.
For cases A and C, a third image appears, resulting in a significant increase
of the image separation.

\subsection{The Non-effect of the Background Matter}

To understand the effect of the background matter, we computed the 
volume density of background
matter along the line of sight, in the vicinity of The Lens. The
resulting density profiles are shown in Figure~6 (solid curves),
where the density is plotted along an interval of comoving length
$128\,\rm Mpc$ centered on The Lens, that is, going from $64\,\rm Mpc$
in front of The Lens to $64\,\rm Mpc$ behind it. Except for
cases E and M, The Lens was always located in a dense region. 
This was expected, since our algorithm locates galaxies predominantly in
clusters, where the density of dark matter is large. Only for cases
E and M The Lens is a field galaxy.

When The Lens is located in a cluster, the
large overdensity of background matter increases the effect of lensing.
However, this effect is greatly compensated by the presence of
low-density regions along the line of sight. The dashed lines show the
line-of-sight-averaged background density within a distance of $\pm64$ Mpc
(comoving) from The Lens. The values are not particularly large, the
maximum being $2.89\bar\rho$ for cases B, C, and D (where $\bar\rho$ is
the universal mean density), while for cases A and K, the line-averaged
value is actually {\it less} than the mean value!
This illustrates the point that was made in \S1, and that was also
made by TOG: along any line of sight, there will be several overdense and underdense regions, and their effects will cancel out at least partly.
As Figure~6 shows, the clusters where The Lenses are located tend
to be surrounded by underdense regions (solid curve dipping below the
dotted line). The only cases where the locally-averaged
background density was more than 100\% larger than
the universal background density ($\rho/\bar\rho>2$) are cases B, C, D, and L.
Case L turns out to be one of the few cases for which the effect of
the background matter exceeds the effect of the galaxies, as Figure~4 shows.

\subsection{The Non-effect of the Galaxies}

Since The Lens tends to be located in a dense cluster, we expect to find
many galaxies nearby. However, only in two cases, D and P, a second
galaxy located in the same plane as The Lens was directly hit (see Fig.~2).
The geometric cross-section of galaxies, $\sigma_{\rm geom}=\pi r_{\max}^2$,
is sufficiently small that galaxies located on the same plane do not overlap
much when seen in projection. This point was already made in Paper~II
(see Fig.~10 in that paper). Hence, if the galaxies physically
associated with The Lens (that is, belonging to the same cluster) have
any effect at all, it will usually result from their combined tidal
field. For this effect to be important, two conditions must be
satisfied: (i) The combined mass of these galaxies must be important, and (ii)
their distribution around The Lens must be asymmetric.

To identify the cluster in which The Lens belong, we simply find
all galaxies located inside a sphere of comoving radius
$R=5\,\rm Mpc$ centered around The Lens. This value of $5\,\rm Mpc$
corresponds roughly to the half-width of the density peaks seen
in Figure~6, and can therefore be used as a fiducial cluster radius.
The galaxies found inside that sphere are considered to be members of
the same cluster as The Lens. Notice that even though The Lens is located
in the center of the sphere used to identify cluster members, it is not
necessarily located in the center of the cluster itself.
In Table~2, we give, for each case, the number of galaxies $N$ in the cluster,
the mass $M_{\rm Lens}$ of The Lens, the total mass
$M_{\rm total}$ of the galaxies in the cluster, the ratio
$M_{\rm Lens}/M_{\rm total}$, and the rank of The Lens, where rank $n$ means
that The Lens is the $n^{\rm th}$ most massive galaxy in the cluster.
Except of cases B, C, D, L, and M, the mass in The Lens is a small fraction
of the total cluster mass, less than 10\%. Of course, The Lens is directly
hit by the beam, while other galaxies contribute only by their tidal
fields. Consequently, even though The Lens does not dominate the mass
of the cluster, it is the most dominant lensing component.

The Lens is the most massive galaxy in the cluster for 5 cases,
and among the five most massive galaxies for 13 cases out of 16, the
exceptions being cases N, O, and P. Figure~4 shows that these three cases
are among the ones for which the effect of galaxies is the largest, with
an increase in magnification larger than 10\% for case P. Notice the large
distortion of the image seen in Figure~3 for that particular case,
when galaxies are added.

The non-effect of galaxies near The Lens results primarily from the
cancellation of their tidal fields. Since The Lens is usually one of the most
massive galaxy in the cluster, it tends to be located near the cluster
center, a consequence of mass segregation.\footnote{In our algorithm,
mass segregation is not achieved dynamically, but is a mere consequence of
morphological segregation. Galaxies located near the center of clusters are
predominantly early-types, and they tend to be more massive than spirals
(see larger values of $v_0$ in Table~1).} In that case, the other
galaxies in the cluster surround The Lens, leading to a substantial
cancellation of their tidal fields.

\section{VARYING THE LENS DENSITY PROFILE}

Our ray-tracing algorithm treats galaxies as nonsingular isothermal spheres.
Each galaxy is described by three parameters, the core radius $r_c$,
maximum radius $r_{\max}$, and rotation velocity $v$, which are related to
the galaxy luminosity and morphological type according to equations (8)--(10).
In this section, we investigate the dependence of our results upon this
particular choice, by considering other density profiles.

\subsection{Singular Isothermal Sphere}

The simplest profile to consider is the singular isothermal sphere.
To obtain this profile, we simply set $r_c=0$ in equation~(7). Hence, we are
essentially replacing every galaxy in our simulation by a singular
isothermal sphere with the same rotation velocity and maximum radius. 

\subsection{NFW Profile and Truncated Isothermal Sphere}

We also consider two other density profiles. The first one is the
widely-used Navarro-Frenk-White (NFW) profile (Navarro, Frenk,
\& White 1997), given by
\begin{equation}
\rho(r)={\rho^{\phantom2}_{\rm NFW}
\over(r/r^{\phantom2}_{\rm NFW})(r/r^{\phantom2}_{\rm NFW}+1)^2}\,,
\end{equation}

\noindent where $r_{\rm NFW}$ and $\rho_{\rm NFW}$ are a characteristic
radius and density, respectively. 
The second one is the Truncated Isothermal Sphere (TIS), a model that was
derived by
Shapiro, Iliev, \& Raga (1999) and Iliev \& Shapiro (2001).
These authors computed the minimum-energy solution of the 
isothermal Lane-Emden equation, modified to account for accretion by
cosmological infall. The actual solution can only be obtained numerically,
but is well-approximated by the following expression,
\begin{equation}
\rho(r)=\rho^{\phantom2}_{\rm TIS}
\left({A\over a^2+r^2/r_{\rm TIS}^2}-{B\over b^2+r^2/r_{\rm TIS}^2}\right)\,,
\end{equation}

\noindent where $r^{\phantom2}_{\rm TIS}$ and 
$\rho^{\phantom2}_{\rm TIS}$ are a characteristic
radius and density, respectively, and $A=21.38$, $B=19.81$, $a=3.01$,
$b=3.82$ (Shapiro et al. 1999).  
The lensing properties of the NFW profile and
TIS are described in great details by Martel \& Shapiro (2003,
and references therein). 

The parameters $r^{\phantom2}_{\rm NFW}$,
$\rho^{\phantom2}_{\rm NFW}$, $r^{\phantom2}_{\rm TIS}$ and 
$\rho^{\phantom2}_{\rm TIS}$ are not free parameters, but functions of the
halo mass, redshift, and the cosmological model
(Navarro et al. 1997; Shapiro et al. 1999; Eke, Navarro, \& Steinmetz 2001;
Iliev \& Shapiro 2001).
However, these models are based on theoretical assumptions that
are inconsistent with the approach we used to generate the distributions
of galaxies (see Paper I). Consequently, we will treat the
characteristic radii and densities as free
parameters, thus treating the NFW profile and the TIS as mere fitting 
formulae. The problem is then the following: considering a galaxy
represented as a nonsingular isothermal sphere described by
equation (7)--(10), we need to derive expressions relating the
characteristic radius and density to $r_c$, $r_{\max}$, and $v$, such
that the same galaxy can be represented as either a NFW profile or a TIS.
Since each profile has two parameters, we need to impose
two constraints. First, the characteristic radii are
related to the tidal radius $r_{200}$ by 
$r^{\phantom2}_{\rm NFW}=r_{200}/C$ and 
$r^{\phantom2}_{\rm TIS}=r_{200}/\zeta_{200}$, where $C$ and $\zeta_{200}$ are
functions of the halo mass, redshift, and the cosmological model 
(for $\zeta_{200}$, the dependence is weak). We assume that the tidal
radius $r_{200}$ is equal to the maximum radius $r_{\max}$.\footnote{The TIS
actually has a truncation radius $r_t\approx1.2r_{200}$. Setting that
truncation radius equal to $r_{\max}$ would be an alternative.}
For the NFW
profile, we use a fixed value $C=9$, which is the value predicted by
the NFW model for a halo of mass $M\sim2\times10^{12}M_\odot$ at
redshift $z\sim0.6$ in a $\Lambda$CDM universe. This value is therefore
reasonable for the galaxies we have identified as ``The Lens.''
For the TIS profile, we use the canonical value $\zeta_{200}=24.2$, which is
correct for all halos in an Einstein-de~Sitter universe. As
Iliev \& Shapiro (2001)
showed, the dependence of $\zeta_{200}$ on the cosmological
parameters is weak, so using this value for a $\Lambda$CDM
universe is not a bad
approximation.

This fixes the values of the characteristic radii 
$r^{\phantom2}_{\rm NFW}$ and $r^{\phantom2}_{\rm TIS}$. Our second
assumption is that
the rotation velocity $v_{\rm rot}$ at radius $r=r_{\rm max}$ is the same
for all profiles.  This ensures that the mass of each galaxy remains fixed,
no matter what density profile is considered.
The rotation velocities are given by
\begin{eqnarray}
v_{\rm rot}^2(r)&=&{4\pi G\rho^{\phantom2}_{\rm NFW}r_{\rm NFW}^3\over r}
\left[\ln(1+r/r^{\phantom2}_{\rm NFW})
-{r/r^{\phantom2}_{\rm NFW}\over1+r/r^{\phantom2}_{\rm NFW}}\right]\,,
\quad{\rm(NFW)}\,;\\
v_{\rm rot}^2(r)&\!\!\!\!=\!\!\!\!&4\pi G\rho^{\phantom2}_{\rm TIS}r_{\rm TIS}^2
\bigg[A-B-
{aAr^{\phantom2}_{\rm TIS}\over r}\arctan{r\over ar^{\phantom2}_{\rm TIS}}
+{bBr^{\phantom2}_{\rm TIS}\over r}\arctan{r\over br^{\phantom2}_{\rm TIS}}
\bigg]\,,\quad{\rm(TIS)}\,;
\end{eqnarray}

\noindent (Chiba \& Takahashi 2002; Martel \& Shapiro 2003).
We set $r=r_{\max}$, equate these expressions to $v^2$, and solve
for the characteristic densities. We get
\begin{eqnarray}
\rho^{\phantom2}_{\rm NFW}&=&{Cv^2\over4\pi Gr_{\rm NFW}^2}
\left[\ln(1+C)-{C\over1+C}\right]^{-1}\,,\\
\rho^{\phantom2}_{\rm TIS}&=&{v^2\over4\pi Gr_{\rm TIS}^2}
\left[A-B
-{aA\over\zeta_{200}}\arctan{\zeta_{200}\over a}
+{bB\over\zeta_{200}}\arctan{\zeta_{200}\over b}
\right]^{-1}\,.
\end{eqnarray}

Once the parameters $r^{\phantom2}_{\rm NFW}$, $\rho^{\phantom2}_{\rm NFW}$,
$r^{\phantom2}_{\rm TIS}$, and $\rho^{\phantom2}_{\rm TIS}$ are
determined, it is straightforward to implement these density profiles into
the ray-tracing algorithm, using the expressions given by Chiba \& Takahashi
(2002) and Martel \& Shapiro
(2003) for the deflection angle.

We selected the particular cases B, C, J, and P, and
redid all experiments, using the various density profiles. 
Figure~7 shows the ratio $\mu/\mu_{\rm Lens}$ for these experiments.
The results are consistent with the ones shown in Figure~4.
The effect of the background matter is of order 1\% at most, the
effect of the galaxies is of order 10\% at most, and the effect of
the other lens planes can be a factor of several. The only
exception is case P with the TIS profile, where the effect of the
galaxies (solid symbols) is of order 40\%.  

\section{SUMMARY AND CONCLUSION}

We have performed a series of ray-tracing experiments using a multiple 
lens-plane algorithm. We selected 16 cases of strong lensing.
By following the trajectory of the beam from the source to the observer,
we were able to determine which galaxies along the line of sight were
directly hit by the beam. The galaxy that produced the largest value of
the convergence $\kappa$ was identified as The Lens, the galaxy
primarily responsible for strong lensing. Our objective
was then to study the effects and relative importance of the various components
along the line of sight, including (i) The Lens itself,
(ii) the galaxies located at the same redshift as The Lens, and possibly
associated with it (i.e. in the same cluster), (iii) the background
matter located at the same redshift as The Lens, and (iv) the matter located
at different redshifts.

With the exception of case M, The Lens was always located in the
redshift interval $0.3\leq z\leq1.1$, where the lensing weight
$w(z)=1/\sigma_{\rm crit}$ is large (see Fig.~2). With the exception of cases
E and M, The Lens was always in a region where the background density
is large (see Fig.~6).
This was expected; strong lensing is caused by massive elliptical and
S0 galaxies, which tend to be located in dense environments according to the
morphology-density relation. In 13 cases out of 16, The Lens was among
the 5 most massive galaxies within a comoving distance of $5\,\rm Mpc$.

The effect of the galaxies and background matter associated with
The Lens are rather small. The magnifications and image separations
vary by $\sim1\%$ when the background matter is added, and $\sim10\%$
when the galaxies are added, for all cases. In some cases (B, K, and P),
a deformation of the image is clearly visible (Fig.~3).
Adding the other lens planes produces effects that range from insignificant
to spectacular. Insignificant effects occur when The Lens is the only large
mass concentration hit by the beam, while
spectacular effects occur when several
galaxies are hit by the beam, including galaxies as massive or more massive
than The Lens.  
The presence of additional galaxies along the line of sight can
create additional images (cases A, M, and O), turn arcs into rings
(cases D and N), and cause large increases in magnification, up to
factors of several. The effect of the other lens planes on the
image separation is less spectacular, but the accumulated
effect is still larger than the effect
of the galaxies and background matter associated with The Lens.

Based on these results, we can summarize the effect of the various
components along the line of sight, relative to the effect of The Lens itself,
as follows:

\begin{itemize}

\item Background matter located near The Lens: a few percent 
(difference between squares and triangles in Figs.~4 and 8).

\item Galaxies near The Lens: several percents, up to $10-15\%$
(cases K and P) (differences between open and filled
symbols in Figs.~4 and 8).

\item Galaxies and background matter on
other planes: factor of several (difference
between filled squares and asterisks in Figs.~4 and 8).

\end{itemize}

We conclude that environmental effects usually play a minor role in strong
gravitational lensing. The large magnification associated with strong
lensing results primarily from a single, massive galaxy (The Lens),
or from the random alignment of several physically unassociated
galaxies at different distances. It is important to insist that this
conclusion is restricted to strong lensing, and cannot be
generalized to weak lensing. We showed that environmental effects
can be of order 10\%. If, say, a massive galaxy, acting as a lens,
increases the brightness of a distant source by 10\%, and the nearby
galaxies and background matter increase it by an additional 10\%, the
net effect is increased by
a factor of 2. Hence, environmental effects can be very important
for weak lensing, but are usually not important for strong lensing.
We have experimented with various density
profiles of galaxies, and reached the same conclusion.
In all these experiments,
{\it we have not found one single case for which the nearby
background matter or nearby galaxies make any significant difference.}

One could debate the statistical significance of our results. We have
considered a subsample of 16 images. Had we considered a larger subsample,
we might have found a case for which environmental effects are important.
For instance, we might find a cluster containing two massive galaxies
that happen to be aligned with the source, so that the beam hits both
galaxies near their center. In this case, one galaxy would be identified as
The Lens, and the effect of the other galaxy would be very important.
Such cases must be very rare, though. The
fact that a subsample of 16 cases has not turned up one single case
for which environmental effects are important, in spite of
the fact that our sample only included cases of very strong lensing, 
suggests that any case
for which such effects are important must be atypical. In this, we
reach the same conclusion as TOG. Note that this results justifies
{\it a posteriori\/} our approach of using a biased subsample of 16 cases.
If we had found, say, 10 or 11 cases for which environmental effects were
very strong, the whole approach would have broken down, but no such
cases were found. 

\acknowledgments

All calculations were performed at the Texas 
Advanced Computing Center, University of Texas.
This work was supported by NASA ATP Grants NAG5-10825, NAG5-10826,
and NAG5-13271. HM thanks the Canada Research Chair program for support.
 
%

\clearpage

\begin{deluxetable}{lcccc}
\footnotesize
\tablecaption{Galaxy Parameters}
\tablewidth{0pt}
\tablehead{
\colhead{Type} & \colhead{$r_0$ ($h^{-1}\rm kpc$)} & 
\colhead{$r_{\max0}$ ($h^{-1}\rm kpc$)} & 
\colhead{$v_0$ ($\rm km\,s^{-1}$)} & \colhead{$\gamma$}
}
\startdata
Elliptical & 0.1 & 30 & 390 & 0.250 \\
S0         & 0.1 & 30 & 357 & 0.250 \\
Spiral     & 1.0 & 30 & 190 & 0.381 \\
\enddata
\end{deluxetable}

\begin{deluxetable}{lccccc}
\footnotesize
\tablecaption{Galaxies near The Lens}
\tablewidth{0pt}
\tablehead{
\colhead{Case} & \colhead{$N(<5\,{\rm Mpc})$} & 
\colhead{$M_{\rm Lens} [10^{12}M_\odot]$} & 
\colhead{$M_{\rm total} [10^{12}M_\odot]$} & 
\colhead{$M_{\rm Lens}/M_{\rm total}$} & \colhead{Rank}
}
\startdata
A       &  70 & 1.15 &  16.11 & 0.071 &  5 \\
B, C, D &  44 & 3.59 &  21.81 & 0.164 &  2 \\
E       &  82 & 2.40 &  27.34 & 0.088 &  3 \\
F       & 101 & 1.31 &  24.69 & 0.053 &  3 \\
G       & 101 & 1.92 &  29.29 & 0.066 &  2 \\
H, I, J & 318 & 4.64 & 104.76 & 0.044 &  1 \\
K       &  36 & 3.05 &  12.96 & 0.235 &  1 \\
L       & 118 & 1.89 &  30.04 & 0.063 &  1 \\
M       &  40 & 2.28 &  10.98 & 0.208 &  2 \\
N       & 254 & 1.10 &  87.10 & 0.013 & 22 \\
O       & 164 & 0.87 &  38.87 & 0.022 & 14 \\
P       & 148 & 0.62 &  35.17 & 0.018 & 18 \\
\enddata
\end{deluxetable}

\begin{figure}
\centering
\vspace{-100pt}
\includegraphics[width=5in]{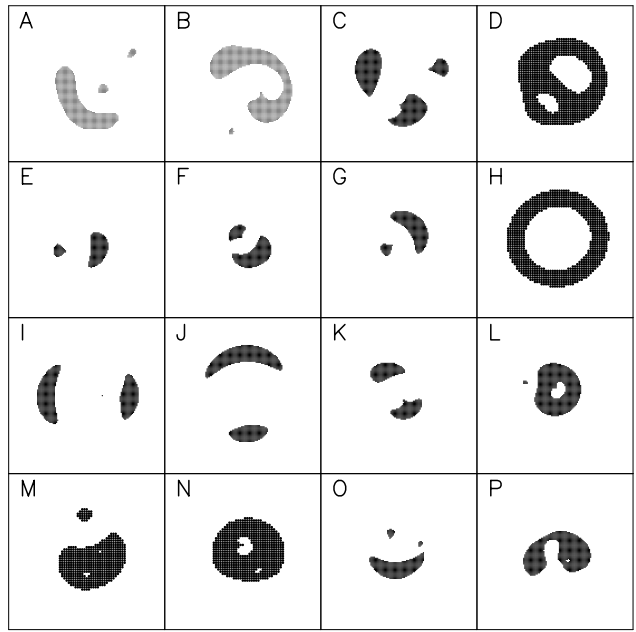}
\vspace{0pt}
\caption{All-plane lensing 
images of the 16 particular cases we have selected for further
analysis. These cases include double images (B, E, F, G, I, J, and K),
triple images (A, C, and O), Einstein ring (H), rings with secondary
image (L and M), double rings (D and N), and peculiar image (P). All
panels ar $9''\times9''$ in size.}
\end{figure}

\begin{figure}
\centering
\vspace{-50pt}
\includegraphics[width=6.5in]{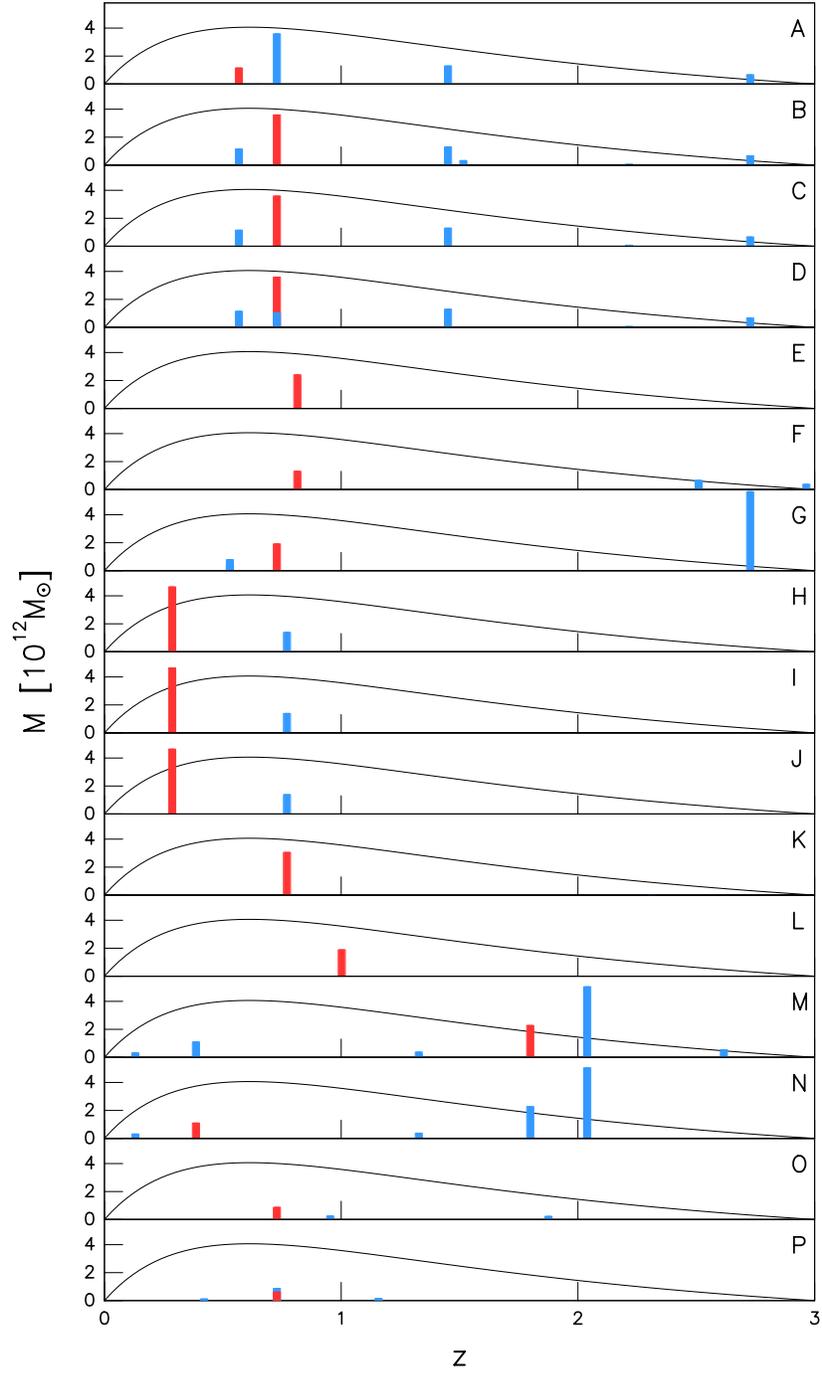}
\vspace{0pt}
\caption{Bar diagram showing the mass of the galaxies that are
hit by the source's central ray, versus redshift. The red bar indicates the
galaxy that was identified as The Lens. Blue bars indicate the
other galaxies. The solid curves show the quantity $1/\sigma_{\rm crit}$,
in arbitrary units.}
\end{figure}

\begin{figure}
\centering
\vspace{0pt}
\includegraphics[width=7in]{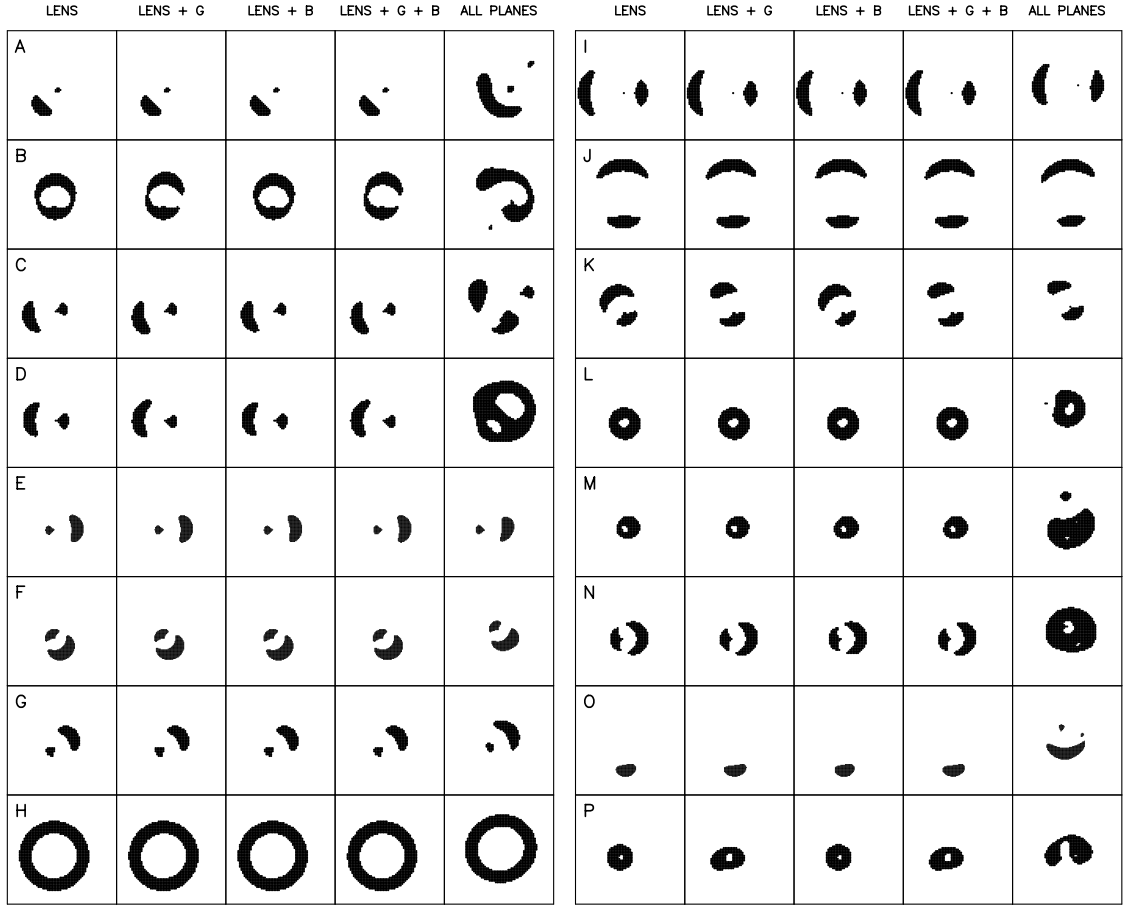}
\vspace{-200pt}
\caption{Images resulting from lensing by various components. Each
row corresponds to a particular case, labeled A through P. First column:
Lens only; second column: Lens + galaxies on same plane; third column:
Lens + background matter on same plane; fourth column: Lens + galaxies and
background matter on same plane; fifth column: all planes included.}
\end{figure}

\begin{figure}
\centering
\vspace{0pt}
\includegraphics[width=7in]{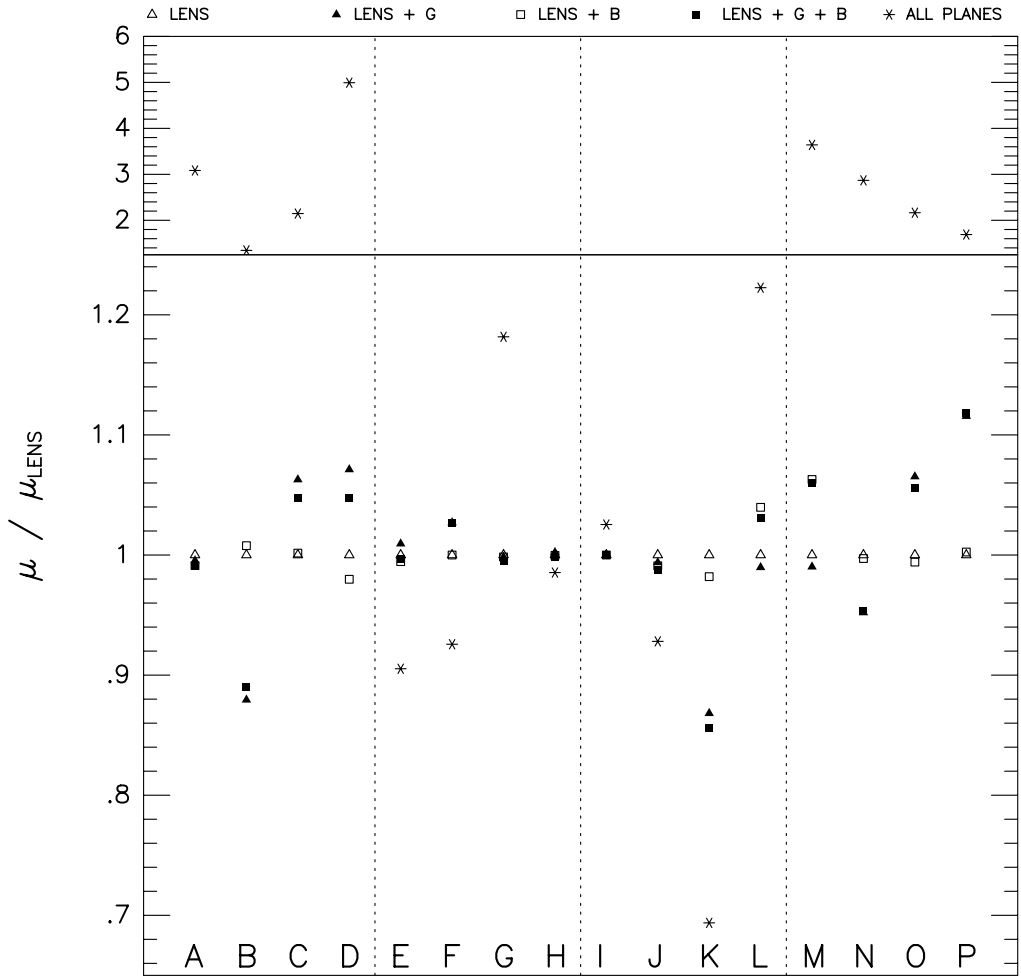}
\vspace{-150pt}
\caption{Ratio $\mu/\mu_{\rm Lens}$, where $\mu$ is the magnification
and $\mu_{\rm Lens}$ is the value of $\mu$ for the Lens case.
The various symbols correspond to various types of experiments, as labeled.
The various cases are identified by the letters at the bottom of the figure.}
\end{figure}

\begin{figure}
\vspace{0pt}
\hspace{50pt}
\includegraphics[width=7in]{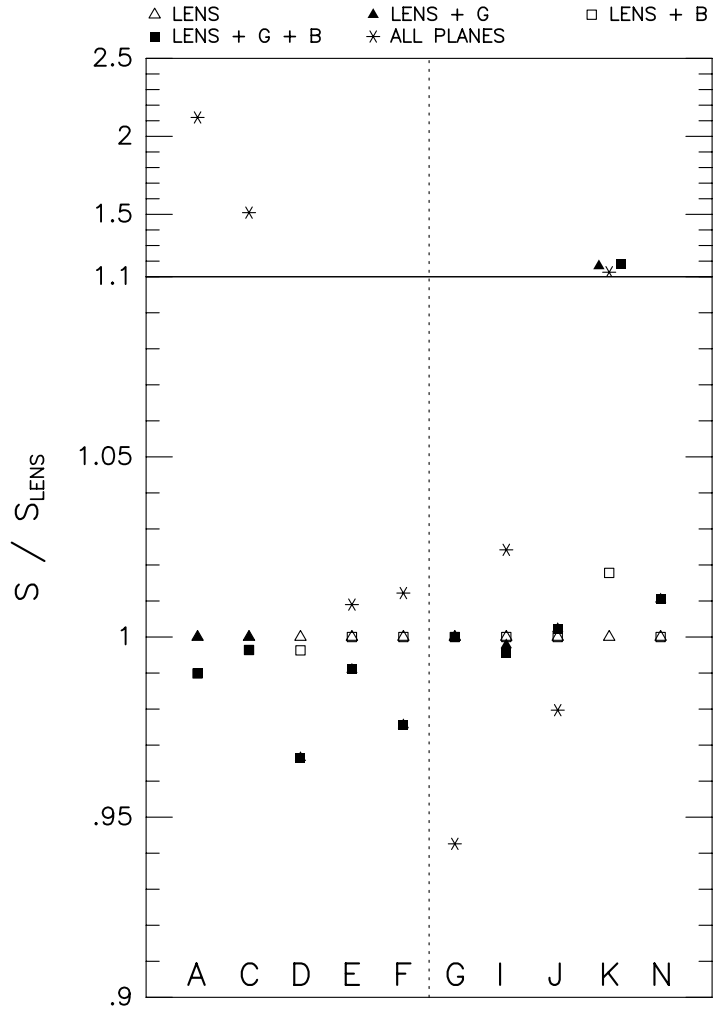}
\vspace{-130pt}
\caption{Ratio $s/s_{\rm Lens}$, where $s$ is the image separation
and $s_{\rm Lens}$ is the value of $s$ for the Lens case.
The various symbols correspond to various types of experiments, as labeled.
The various cases are identified by the letters at the bottom of the figure
(Cases B, H, L, M, O, and P do not produce multiple images). For case
K, we have displaced the filled symbols sideway for clarity.}
\end{figure}

\begin{figure}
\centering
\vspace{-20pt}
\includegraphics[width=7in]{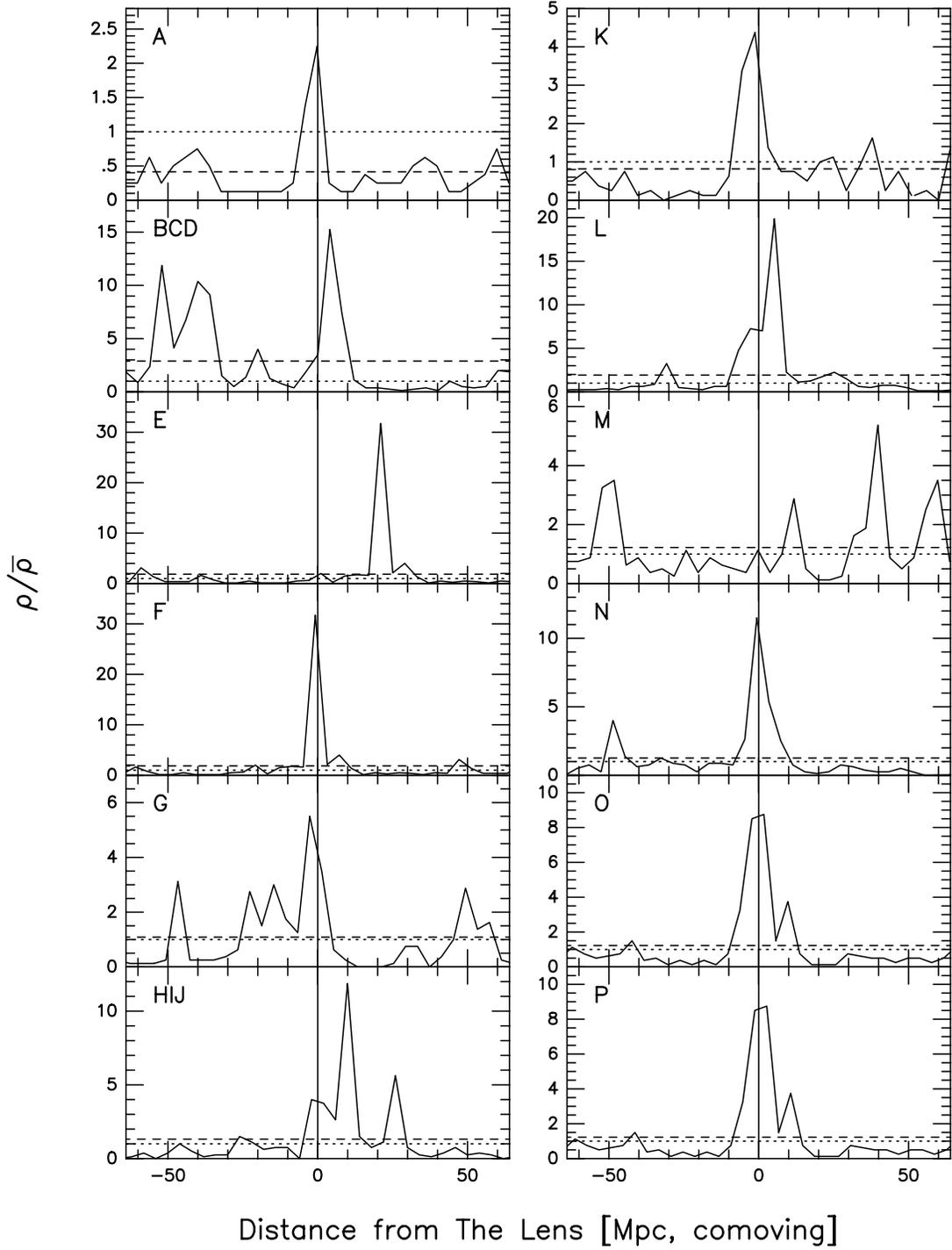}
\vspace{-50pt}
\caption{Solid curves: density profile $\rho$ of background matter along
line of sight, in units of the universal mean density $\bar\rho$ of
the universe, vs. position relative to The Lens. The dotted lines shows 
the universal mean density. The dashed lines show the mean density along the
line of sight. The vertical lines indicate the location of The Lens.
All cases are labeled. Notice that for cases B, C, and D, and for
cases H, I, and J, The Lens was the same galaxy.}
\end{figure}

\begin{figure}
\centering
\vspace{0pt}
\includegraphics[width=7in]{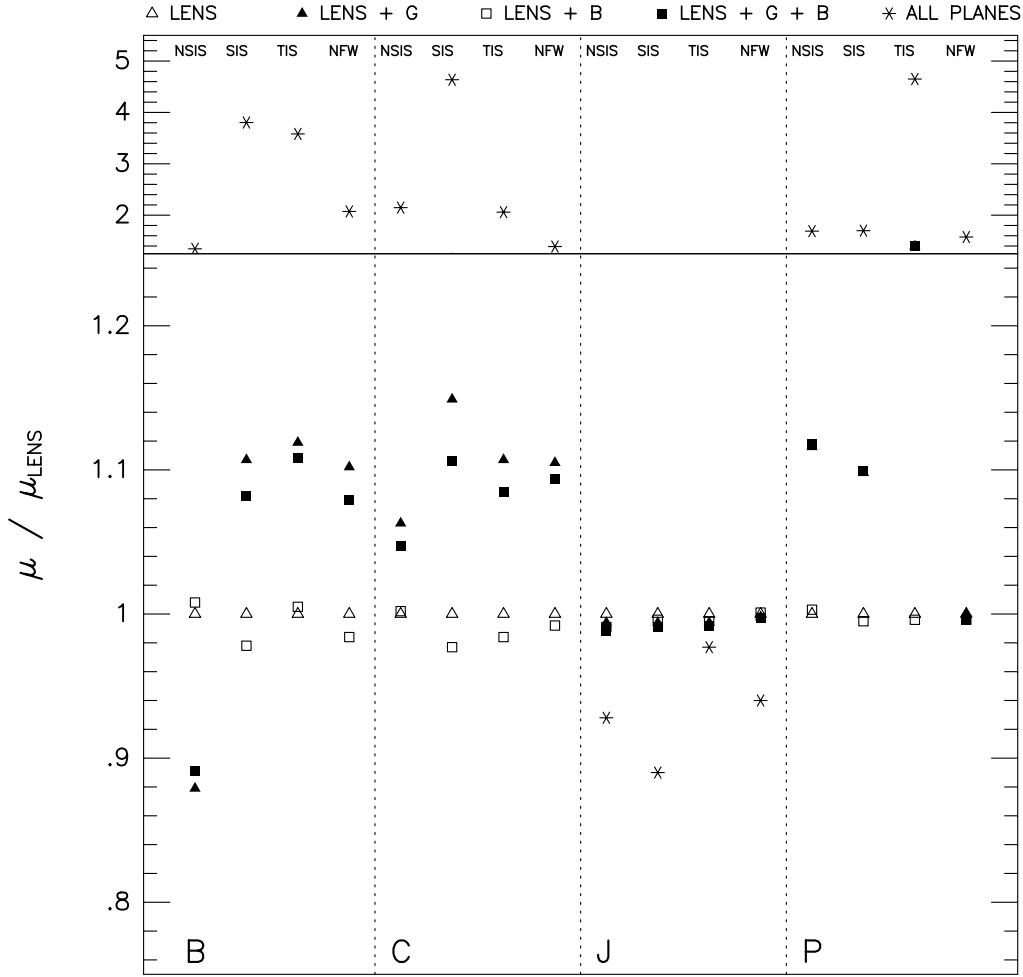}
\vspace{-150pt}
\caption{Ratio $\mu/\mu_{\rm Lens}$, where $\mu$ is the magnification
and $\mu_{\rm Lens}$ is the value of $\mu$ for the Lens case, for cases
B, C, J, and P, with various with various density
profiles for galaxies.
The various symbols correspond to various types of experiments, as labeled.
The various cases are identified by the letters at the bottom of the figure.
The density profiles are indicated at the top of the figure.}
\end{figure}

\end{document}